# MPI同步通信顺序模型死锁静态检测算法

**廖名学，何晓新，范植华**

（中国科学院软件研究所，北京海淀区中关村南四街4号 100080）

**摘 要：** 静态检测 MPI 程序同步通信死锁比较困难，通常需要建立程序模型。顺序模型是其他所有复杂模型的基础。通过一种映射方法将顺序模型转化为字符串集合，将其死锁检测问题转化为等价的多队列字符串匹配问题，从而设计并实现了一种 MPI 同步通信顺序模型的静态死锁检测算法。算法时空复杂度均为 O(n)，这里 n 是模型中消息的总数。算法性能好于通常的环检测方法并能适应动态消息流。

**关键词:** MPI；算法；死锁；静态分析

## Algorithm of Static Deadlock Detection in MPI Synchronization Communication Sequential Model

**Liao Ming-Xue, He Xiao-Xin, Fan Zhi-Hua**

(Institute of Software, the Chinese Academy of Sciences, 4#4th South Street, Zhong guan cun, Haidian District, Beijing 100080)

【Abstract】Detecting deadlocks in MPI synchronization communication programs is very difficult and need building program models. All complex models are based on sequential models. The sequential model is mapped into a set of character strings and its deadlock detection problem is translated into an equivalent multi-queue string matching problem. An algorithm is devised and implemented to statically detect deadlocks in sequential models of MPI synchronization communication programs. The time and space complexity of the algorithm is O(n) where n is the amount of message in model. The algorithm is better than usual circle-detection methods and can adapt well to dynamic message stream.

【Key words】MPI; algorithm; deadlock; static analysis

　　死锁是计算机科学中一个重要的研究对象。1968年Dijkstra中描述了有限资源环境中并发进程资源请求过程可能出现的死锁并给出了银行家算法。1971年Coffman给出了死锁产生的4个必要条件以及处理死锁的三种策略：死锁预防，死锁避免和死锁检测与恢复。Mukesh在[1]中认为死锁预防和死锁避免具有局限性，难度很高，主要用在可靠性要求很高的系统中，死锁处理大多采用死锁检测。近几年针对具体程序语言上的死锁研究取得了很多成果，如Christoph von Praun在其博士论文[2]中给出的多线程面向对象程序（主要指Java）同步死锁检测算法等。

　　MPI[3]是分布式内存并行处理计算机上采用Fortran，C和C++语言开发基于消息传递应用系统的标准（其主要实现是MPICH），主要用于大规模并行计算机和集群的高性能运算，但其调试极其困难。目前MPI上的调试很大程度上集中于通信死锁检测，一般都采用动态方法（运行时），比如[4]。

　　本文试图从静态预防上探讨一种新的方法来检测MPI程序中的同步通信死锁问题。静态死锁检测通常需要建立模型，如[5]所建立的工作流模型。顺序模型研究是其他一切复杂模型如循环模型死锁研究的基础，复杂模型的死锁判定最终全部要规约为顺序模型死锁判定；也正是如此，设计性能良好的顺序模型死锁检测算法是非常关键的。本文第1节抽象出MPI同步通信的顺序模型，第2节给出一个线性时空复杂度的死锁检测算法并分析其性能，第3节给出了算法的实现概貌，最后得出结论。

## 1 问题描述

本文考虑的问题针对如下 MPI 程序（真实程序的抽象）：

$Process(machine/node)\, i:\ send(recv)\ Message\ x\ To\ (From)\ Process\ j$     (1)
这里 $, 0 \leq i \neq j < n, n$ 是参与并行计算的节点个数，$x$ 表示一种消息

比如下面运行于三个节点上的 MPI 程序：

| *Process(machine) 0* | *Process(machine) 1* | *Process(machine) 2* | |
|---|---|---|---|
| send Message *a* To *Process 2* | receive Message *b* From *Process 0* | receive Message *c* From *Process 1* | (2) |
| send Message *b* To *Process 1* | send Message *c* To *Process 2* | receive Message *a* From *Process 0* | |

明显地，上面的MPI通信程序存在一个等待环从而导致死锁，而检测环的时间复杂度是 $\Theta(|V|+|E|)$ [6]，其中|V|相当于下文谈到的消息总数，E表示图的边集合，并且还需要建立消息顺序图[1]。于是有如下问题：

是否存在一个比环检测更加高效的算法来判断程序(1)死锁？     (3)

## 2 算法

MPI 程序(1)可以抽象成字符串集合，如(2)可以抽象成 3 个字符串：*ab,bc,ca*。抽象过程是简单的，因此并不需要严格定义。如此抽象之后，MPI 程序(1)的死锁问题就演变为字符串集合多队列匹配问题。

### 2.1 字符串集合上多队列匹配规则及算法

规则 1：如果所有字符串全部为空，那么认为字符串集合匹配成功。
规则 2：如果有某种字符不是正好出现在两个字符串中，那么认为字符串集合是非法的。
规则 3：如果有相同字符出现在两个字符串之首，那么将这两个字符从相应的串中删除。
规则 4：如果前面所有规则都不能应用于字符串集合，那么认为字符串集合匹配失败。

规则 2 在MPI程序中的意义是：点对点通信的消息只可能有唯一一个发送者和唯一一个接收者，因此一种消息必须且只能存在于两个进程中。规则 3 在MPI程序中实际上就表示消息收发过程。规则 4 类似于[7]所说的"水平面原理"，即没有进程能够继续向前推进。从上述匹配规则可知，程序(1)的死锁问题等价于与(1)对应的字符串集合的多队列匹配问题。下面给出匹配算法：

匹配算法:checkDeadlock  算法输入:字符串数组 sss
/*消息(字符)哈希表, 初始为空. 表结构=<消息,匹配器>, 消息是键, 即每种消息都有一个匹配器. 匹配器记录消息(字符)在哪些字符串中出现过.*/
  message-table
  sss-queue/*sss 对列, 初始包括了 sss 中所有字符串.*/
  while(sss-queue 非空)
    next-sequence←sss-queue 出队/*取队首字符串*/
    next-message←next-sequence 下一个字符(消息)
    if(message-table 中没有消息next-message)创建该消息的匹配器并连同该消息加入到消息哈希表
    else 在该消息的匹配器中记录 next-sequence endif
    if(next-message 的匹配器发现该消息出现在两个以上字符串(序列))
        根据规则2断言字符串集合 sss 匹配失败
    else if(next-message 的匹配器发现 next-message 正好出现在两个字符串中)
        根据规则3处理这两个字符串.如果这两个字符串还有未处理字符, 将它们入队 sss-queue
    end-id end-if
  end-while
  if(某个匹配器发现还有消息没有匹配)根据规则4与规则2宣布死锁并给出死锁位置 end-if
end-checkDeadlock

## 2.2 算法性能分析

首先是关于时空复杂度的评估。While 循环的每一次循环都处理一条消息,因此算法时间复杂度为 $O(n)$,$n$ 是消息总数;消息哈希表的大小最多等于消息种类的总数,而消息种类的总数不大于消息总数(同一种消息可能多次出现),因此最坏空间复杂度不超过 $O(n)$,比环检测方式略好。需注意消息哈希表中要存储消息匹配器,根据匹配规则 2,容易设计一种匹配器,使其空间消耗不大于常数 2(限于篇幅,匹配器的具体数据结构未给出);sss 队列最大长度等于 sss 中字符串的数量,这个数量不大于消息总数,故其最坏空间复杂度不超过 $O(n)$。

其次是关于算法动态适应能力的评估。"动态"指算法的输入 sss 是动态的,即可以随时向 sss 中每个字符串的末尾追加字符,这样就形成字符(消息)流。对更为复杂的模型(比如循环模型)所做的模型检测往往需要先将其变换成含有动态消息流的顺序模型,然后进行死锁检测(因为这些模型在某个时刻或者某个条件下往往只能确定部分消息,后续消息是未知的)。因此,顺序模型的死锁检测算法需要能适应动态消息流。基于环检测的算法显然无法适应这种动态性,而我们提出的算法,在 While 循环中合适的地方添加一段更新 sss 的代码即可适应这种动态性(具体实现中,我们通过增加一行代码,用来从另一个实时更新的线程中获取更新内容,当然,何时不会再有更新是可以预先知道的,这是因为程序和模型结构是不变的,变化的是从那些复杂模型中产生顺序模型的策略,由此保证了算法的终止性)。

另外一个问题是,算法是将 MPI 程序中的消息映射为字符后的处理方法。那么如何才能映射到字符?根据 MPI 标准,点对点消息的消息信封为<消息 tag, 消息源 source, 消息目的地 destination, 消息所在的通信子 communicator>。从上面的算法可以看出,映射的一个根本目的是采用 Hash 表存储消息,也就是说只要能够根据消息信封建立哈希表就可以达到映射的目的。根据 MPI 标准,消息信封的四项内容都是整数。如此,可采用一种多重 Hash 方法管理所有消息。

首先,使用消息签名区分开每种消息。我们将消息签名设计为非负整数类型,即每一种 MPI 消息对应一个独一无二的整数。所有消息的签名定义为签名空间 signature-space,我们需要设计算法来管理整个签名空间,即如何为每种消息正确分配一个独一无二的签名。下面给出的是签名空间的数据结构,在此基础上实现签名空间的一个功能 setSignatureFor(Message) 为消息创建和检索签名,其具体过程不再赘述。

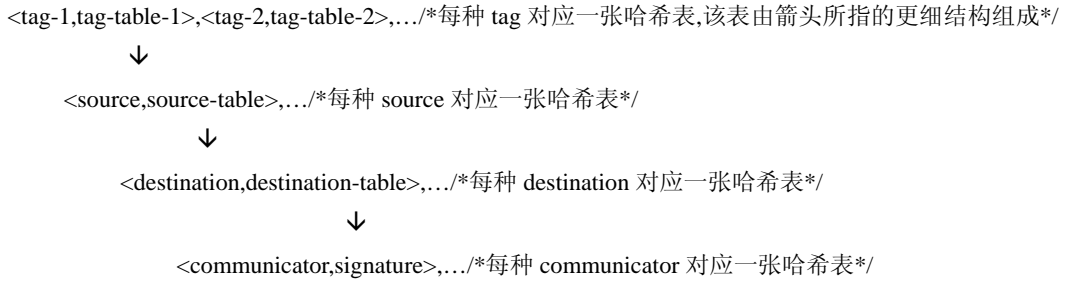

我们可以很容易地设计出签名空间的管理功能,使得该功能每次被调用的运行时间为常数,空间复杂度为 $O(n)$,$n$ 为消息种类总数。

从上述性能分析可知,整个 MPI 顺序模型静态死锁检测算法的时空复杂度为 $O(n)$,$n$ 为消息种类总数,从而回答了第 2 节提出的问题(3)(考虑环检测算法需要输入图的数据结构,本算法的空间开销并不比环检测算法大),并且算法能够很好地适应动态消息流。

## 3 算法实现与应用

本文的算法在我们的 MPI 同步通信死锁静态判定框架中已经全部采用 Java 语言实现并运行良好。该算法实现的类框架如图 1 所示(框架图是整个死锁判定框架的一个子集)。图中,含实心圆的线段表示对象组合关系;实心箭头表示关联关系,空心箭头表示类继承关系。图中没有画出类 Program,这个类用来表示语法分析器解析 MPI 程序所得的结果,其他重要标识符请参考 Java 语言。

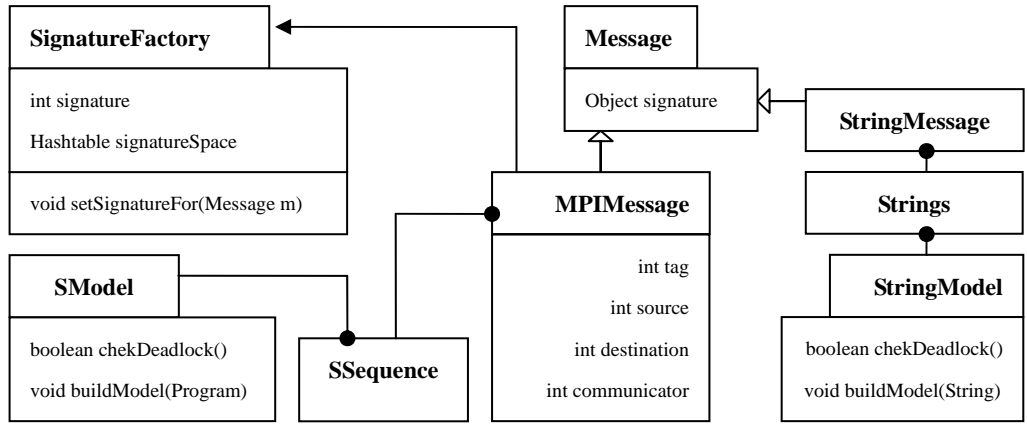

图 1 MPI 同步通信顺序模型静态死锁检测算法实现的类框架图

## 4 结论

本文设计了 MPI 同步通信顺序模型的静态死锁检测算法，该算法时空复杂度为 O(n)，n 为消息总数，比通常的环检测算法(时间复杂度 $\Theta(|V|+|E|)$)更加有效，并且不需要预先建立图的数据结构，为其他复杂模型死锁检测提供了良好基础。更重要的是，该算法还能很好地适应动态消息流，这是基于环检测的死锁检测方法所不可能做到的。目前算法已经全部实现并在实际系统中运行良好，感兴趣的读者可以通过作者 e-mail 索取源代码(本文涉及的源码位于包 mpimodel 下，主要是 SModel 与 SSequence 类，可通过这两个类索引其他相关类)。